\documentclass[trackchanges,twocolumn]{aastex7}
\usepackage{amsmath,textcomp}
\usepackage{enumitem}

\newcommand{\fref}[2]{\hyperref[#1]{\ref*{#1}#2}}

\begin{document}

\title{The Detectability of Lunar-Origin Asteroids in the LSST Era}

\author[orcid=0009-0004-6339-282X, gname=Yixuan, sname=Wu]{Yixuan Wu}
\affiliation{Tsinghua University, Beijing 100084, China}
\email{wu-yx23@mails.tsinghua.edu.cn}

\author[orcid=0000-0003-1097-0521, gname=Yifei, sname=Jiao]{Yifei Jiao}
\affiliation{Tsinghua University, Beijing 100084, China}
\affiliation{Department of Earth and Planetary Science, University of California, Santa Cruz, CA 95064, USA}
\email[show]{jiaoyf.thu@gmail.com}  

\author[orcid=0000-0002-0209-7286, gname=Wen-Yue, sname=Dai]{Wen-Yue Dai}
\affiliation{Tsinghua University, Beijing 100084, China}
\affiliation{National Astronomical Observatory of Japan, 2-21-1 Osawa, Mitaka, Tokyo 181-8588, Japan}
\email{wydai.thu@gmail.com}  

\author[orcid=0000-0003-1215-4130, gname=Yukun, sname=Huang]{Yukun Huang}
\affiliation{National Astronomical Observatory of Japan, 2-21-1 Osawa, Mitaka, Tokyo 181-8588, Japan}
\email{yhuang.astro@gmail.com}  

\author[orcid=0009-0008-0393-2678, gname=Zihan, sname=Liu]{Zihan Liu}
\affiliation{Tsinghua University, Beijing 100084, China}
\email{805219748@qq.com}  

\author[orcid=0000-0002-8025-9113, gname=Bin, sname=Cheng]{Bin Cheng}
\affiliation{Tsinghua University, Beijing 100084, China}
\email[show]{bincheng@tsinghua.edu.cn}  

\author[gname=Hexi, sname=Baoyin]{Hexi Baoyin}
\affiliation{Tsinghua University, Beijing 100084, China}
\email{baoyin@tsinghua.edu.cn}  

\author[gname=Junfeng, sname=Li]{Junfeng Li}
\affiliation{Tsinghua University, Beijing 100084, China}
\email{lijunf@tsinghua.edu.cn}

\begin{abstract}
While most near-Earth asteroids (NEAs) are thought to originate from the main belt, recent discoveries have suggested the existence of a lunar-derived NEA population, such as the asteroids Kamo‘oalewa and 2024~PT5. 
These objects may hold key clues to the dynamical evolution of NEAs and the recent impact history of the Earth\textendash Moon system.
However, the population, distribution, and dynamical characteristics of these Lunar-Origin Asteroids (LOAs) remain poorly constrained. 
By combining the lunar ejecta production with N-body orbital simulations of the ejecta, we investigate their orbital evolution in the past millions of years and the current LOA population, revealing their significant potential for detection by future surveys. 
Specifically for the Vera C. Rubin Observatory's upcoming Legacy Survey of Space and Time (LSST), we predict an average detection rate of about 6 LOAs (with $D > 5 \text{\ m}$) per year. 
Additionally, we find that the LOAs tend to approach from sunward and anti-sunward directions, with encounter velocities significantly lower than those of typical NEAs. 
These findings offer valuable insights in guiding targeted ground-based surveys and planetary defense efforts for LOAs in the future.
\end{abstract}



\section{Introduction} 

Near-Earth asteroids (NEAs) are typically thought to have originated from the main asteroid belt between Mars and Jupiter. 
These asteroids dynamically migrated into Earth's vicinity, driven by the Yarkovsky effects, mean motion and secular resonances with Jupiter, and close encounters with terrestrial planets \citep{gladman1997dynamical,granvik2017escape}. 
However, recent studies revealed the existence of a unique group of ten-meter-sized asteroids in near-Earth space that may have originated from lunar ejecta\textemdash debris ejected during impact events on the lunar surface in the past $\sim$100~Myr. 
These so-called Lunar-Origin Asteroids (LOAs) include the prominent case of Earth's quasi-satellite 469219 Kamo‘oalewa (2016~HO3), target of China's Tianwen-2 asteroid sample return mission \citep{zhang2021china}, whose reflectance spectrum shows a strong similarity to lunar silicates \citep{sharkey2021lunar}. More recently, the Earth's temporarily-captured asteroid 2024~PT5 has also been suggested as a candidate due to its lunar-like spectrum and low relative velocity \citep{kareta2025lunar,bolin2025discovery}.

The impact cratering on the Moon provides a robust production mechanism for this population.  
For instance, a single 1-km-diameter object impacting the Moon would produce $\sim10^7$ meter-sized ejecta exceeding the lunar escape velocity of 2.38~km/s. 
Moreover, the potential 2032 impact of asteroid 2024~YR4 on the Moon, with a forecasted 4.3\% probability \citep{wiegert2025potential,jiao2025probing}, presents a remarkable opportunity to witness the fresh production of meter-sized LOAs.
These ejecta could further enter the Earth\textendash Moon space and then become NEAs \citep{gladman1995dynamical}. 
While a large fraction of these lunar-origin asteroids would collide with the Earth as lunar meteorites within the first few million years after the impact event \citep{castro2025lunar}, some LOAs do survive through millions of years of dynamical evolution.

However, fundamental properties of the current LOA population, such as its total number, remain poorly constrained. 
 Special cases such as lunar-source minimoons are unlikely to yield a meaningful sample, because their predicted numbers are small and their rapid motion (often several degrees per day) makes these faint objects extremely difficult to detect \citep{jedicke2025steady}. 
This highlights the need for a comprehensive assessment of the entire LOA population to determine the potential opportunities for detection.
In addition to their scientific value, LOAs may pose a distinct Earth impact risk. 
Therefore, for the dual purposes of discovering new LOA members and informing planetary defense strategies like kinetic deflection \citep{jiao2023optimal,lee2025investigation}, it is crucial to investigate their dynamical behavior during Earth flybys, which may differ from that of typical NEAs from the main belt.

In this work, we combine the lunar ejecta production and N-body simulations incorporating the Yarkovsky effect \citep{vokrouhlicky2015yarkovsky} to systematically investigate the orbital evolution of $D > 5 \text{\ m}$ lunar ejecta over a 100-Myr timescale, covering the dynamical lifetimes of most LOAs \citep{jiao2024asteroid}. 
Focusing on the Earth flyby events of potential LOAs, our primary analysis assesses their detectability by ground-based optical surveys, in particular the forthcoming Vera C. Rubin Observatory's Legacy Survey of Space and Time (LSST) \citep{ivezic2019lsst}. 
Our analysis of their orbital evolutions further provides insights for both distinguishing this unique population from the broader NEAs and enhancing early-warning capabilities for planetary defense.

\setcounter{footnote}{0} 

\section{Methods} 

\subsection{Lunar ejecta production model}\label{sec:2.1}

\begin{figure*}[ht!]
    \centering
\includegraphics[width=0.6\linewidth]{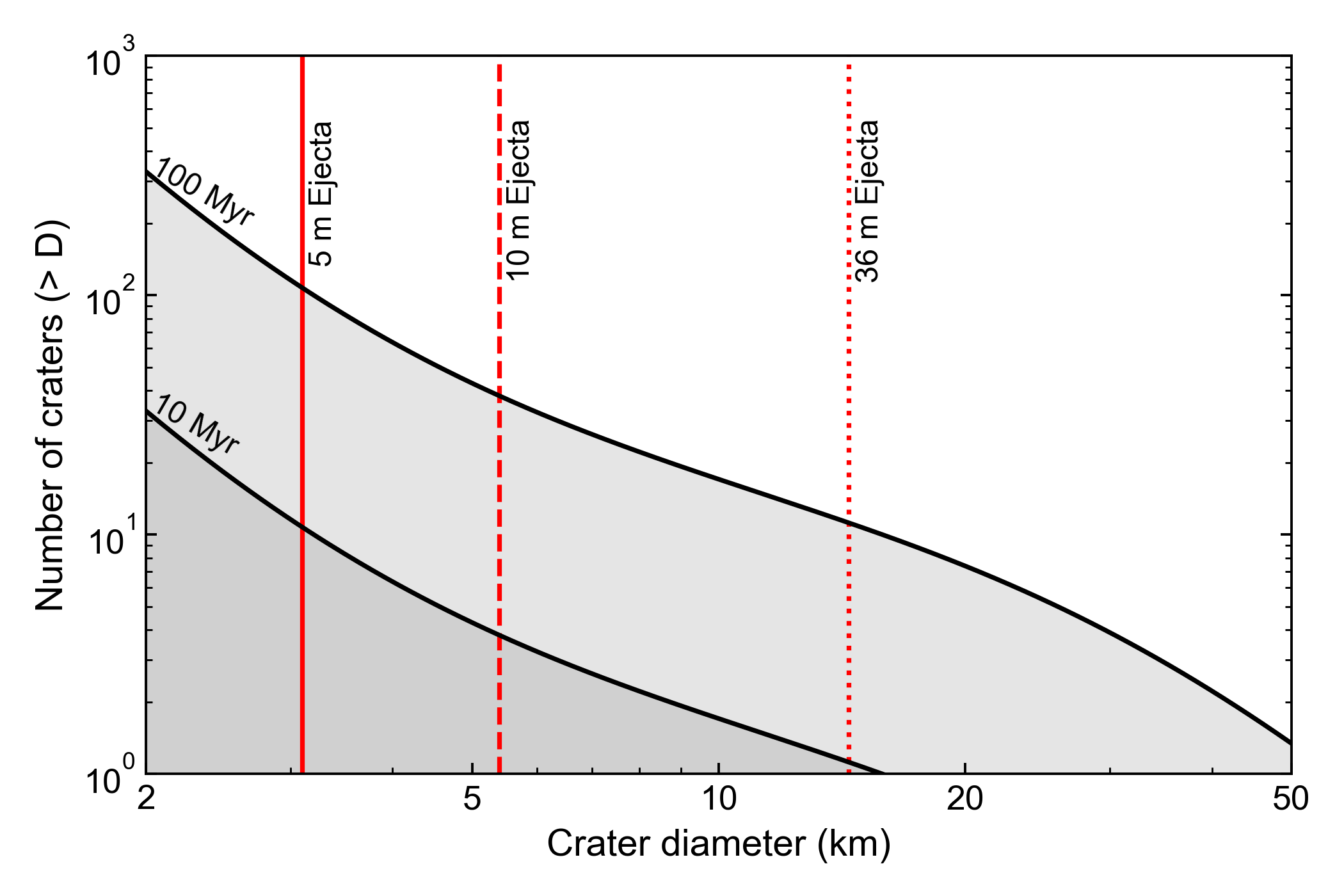} 
\caption{\textbf{ Cumulative size distribution of lunar craters.}The black curves illustrate the cumulative size distribution of lunar craters over the past 10 Myr and 100 Myr, calculated from the background impactor flux.  The \textit{x}-axis scale of the red vertical lines indicates the minimum crater diameter necessary to launch lunar escape ejecta of certain sizes. For instance, producing a Kamo‘oalewa-sized (36 m) fragment requires a crater of at least $\sim$14.4~km in diameter.
\label{fig:1}}
\end{figure*}

To model the production of lunar ejecta, we begin with the lunar cratering history. 
For the entire lunar surface, \citet{neukum2001cratering} and \citet{ivanov2001mars} have studied the size-frequency distribution (SFD) of lunar craters, as shown in Figure \ref{fig:1}. 
For simplicity, we employ a simple power-law approximation of the cratering model. 

In this simplified model, the cumulative number of craters larger than $D_c$ (in km) and younger than the age $t$ (in Myr) is then given as
\begin{equation}
 N_{\text{c}}(>D_{\text{c}},<t)=N_0(t)D_{\text{c}}^{-b}~, \label{eq:1}
\end{equation}
where $b=1.833$ and $N_0(t)=14.473t$.
We first define a ``background'' model from this equation by setting $N_c = 1$ (one expected crater), which gives the statistically expected maximum crater diameter over the past $t$~Myr. 
For $t=10~\text{Myr}$, it predicts a largest crater of ~14.4~km.

However, this ``background'' model, based on a long-term average flux, fails to account for the 22-km Giordano~Bruno crater, which formed within the last 1--10~Myr \citep{morota2009formation}. 
To capture such rare large craters, we introduce a ``real'' model with $N_c = 0.5$, which yields a diameter threshold of ~22-km for craters in the past 10~Myr. The ``real'' model is consistent with Giordano~Bruno and may better match the real LOA population, while the ``background'' model provides a conservative baseline.

Following this cratering history, we further investigate the production of escaping ejecta in three steps.
First, we connect each crater to its impactor using the scaling law from \citep{singer2020lunar} as
\begin{align}
&D_\text{tr}=\begin{cases}1.15D_{\text{c}}^{0.885} & \text{if~}D_c\geq11\text{~km}\\0.83D_{\text{c}}&\text{if~}D_{\text{c}}<11\text{~km}\end{cases}~,\\
&D_{\text{i}}=1.004D_\text{tr}^{1.275}\left[g/\left(v\sin\theta\right)^2\right]^{0.275}\ ,
\end{align}
where $D_\text{tr}$ is the transient crater diameter, and $D_\text{i}$ is the impactor diameter. 
The other parameters denote the lunar gravitational acceleration $g$, the impact velocity $v$, and the impact angle $\theta$ (measured from the ground plane). 
Here, all units are in $\text{km}$ and $\text{s}$.

Then, we calculate the number of escaping ejecta produced by this impactor. 
Based on the Smoothed Particle Hydrodynamics (SPH) simulations by \citet{jiao2024asteroid}, the minimum impactor diameter ($D_\text{{i,min}}$) required to produce escaped ejecta with diameter $> D$ can be estimated as 
\begin{align}
D_{\text{i,min}}=D/0.035~.
\end{align}
The total escape mass have been suggested to be proportional to the impactor mass \citep{artemieva2008numerical}, while the debris size-frequency distribution is typically following a power-law relation \citep{jiao2024asteroid}.
The cumulative number of escaped ejecta with diameter $> D$ is then given by
\begin{align}
N_{\text{e}}(> D)=kD^{-p}D_{\text{i}}^3~,  \label{eq:5}
\end{align}
where $k=5.484\times 10^{-4}$ and $p=3.5$ are determined by SPH simulations \citep{jiao2024asteroid}.
As a reference, there would be about two 10-km lunar craters formed in the past 10 Myr, with each producing about 15,000 ejecta ($>$~5~m) that can escape the lunar gravity.

Finally, we determine the launch velocities of these fragments. The velocity magnitude ($v$) of lunar ejecta can be described by a power-law distribution \citep{jiao2024asteroid}. 
For escaping ejecta ($D > 5 \text{\ m}$), their cumulative frequency distribution is
\begin{align}
&N_{\text{e}}(> v) \propto v^{\alpha} \quad  \text{for } v_{\min} \leq v \leq v_{\max}~,  \label{eq:6}
\end{align}
where $\alpha = -4.0$, $v_\text{min} = 2.38~ \text{km/s}$ is the lunar surface escape velocity, and $v_\text{max} = 6.00~ \text{km/s}$.
The launch direction is then defined by a random azimuth angle and an assumed polar angle of 45° from the surface normal.

\subsection{N-body simulations}

The key assumption of our orbital simulations is that the production and evolution of lunar ejecta are in a steady state. 
This is justified because the impactor flux in the inner Solar System has been relatively constant over the last 3 Gyr \citep{neukum2001cratering}, and the dynamics of ejecta are insensitive to long-term variations in Earth's orbit \citep{castro2025sensitivity}. 
This crucial steady-state condition allows us to equate the survival time of a particle in our forward-in-time simulations with the potential formation age of a present-day LOA.

We perform a suite of N-body simulations using the open-source code REBOUND \citep{rein2012rebound}. 
The full ensemble consists of 200 individual simulations, tracing the evolution of 20,000 test particles. 
To systematically sample the orbital geometry, each of the 200 simulations is assigned a unique, randomly chosen lunar phase. 
Within any given simulation, 100 massless particles are launched simultaneously from random locations on the lunar surface. 
The initial conditions for each particle\textemdash velocity and diameter\textemdash are drawn from power-law distributions (Eqs.~\eqref{eq:5}--\eqref{eq:6}), and the launch angle is fixed at 45$^\circ$ following the Z-model \citep{maxwell1977simple}.

The N-body simulations are conducted following the steps below.
First, we simulate the evolution of the launched particles that escape from the Moon for 100 years under the gravity of the Sun, the eight planets, and the Moon.
By the end of this phase, most survival particles have reached heliocentric orbits. 
The simulation then enters a heliocentric phase where the Earth\textendash Moon system is represented by its barycenter. 
In this phase, the particles are simulated for a time span of 100~Myr, or until they hit a planet or the Sun, or their aphelion exceeds 6~AU (beyond Jupiter's orbit).

To account for the Yarkovsky effect \citep{vokrouhlicky2015yarkovsky} in our long-term simulations, we apply an acceleration to each particle following the model of \citet{veras2019speeding}. 
The acceleration magnitude is determined by the particle's diameter and heliocentric distance, with its direction (inward or outward) initially randomized. 
This leads to a semi-major axis drift rate of up to $\sim 10^{-2}~{\rm AU\,Myr^{-1}}$ for 5-m objects, consistent with calculations for larger objects like 2016~HO3 \citep{hu2023peculiar}, which yielded $(-4.6\pm1.9)\times10^{-3}~{\rm AU\,Myr^{-1}}$.
The solar radiation pressure can be neglected, because it is more than one order of magnitude weaker than the Yarkovsky effect for asteroids with $D > 5 \text{\ m}$ \citep{deo2017yarkovsky,burns1979radiation}.

We record the heliocentric orbits of all survival particles every 10,000 years, with the recording cadence increased to every 0.03 years during Earth close encounters (distance $<$ 0.05~AU) to resolve their detailed dynamics. 
To identify detectable events for LOAs during their Earth flybys, the apparent magnitude $V$ of the asteroids is calculated at each output step using the asteroid photometric model of \citet{bowell1989application}
\begin{align}
&V = H + 5\lg(d_{\text{S}} \cdot d_{\text{E}}) - 2.5\lg q~,\\
&H = 15.6176 - 5 \lg D - 2.5 \lg p_{\rm v}~,
\end{align}
where $d_{\text{S}}$ and $d_{\text{E}}$ represent their distances from the Sun and Earth, respectively. 
Here, $ q$ is a function that depends on the Earth-asteroid-Sun angle, and $D$ is the asteroid's diameter.
To match the physical properties of 2016~HO3, we set the geometric albedo $p_{\rm v}$ to $0.20$ \citep{sharkey2021lunar}.

\section{Results and Discussions } 

\begin{figure*}[ht!]
    \centering
    \includegraphics[width=0.75\linewidth]{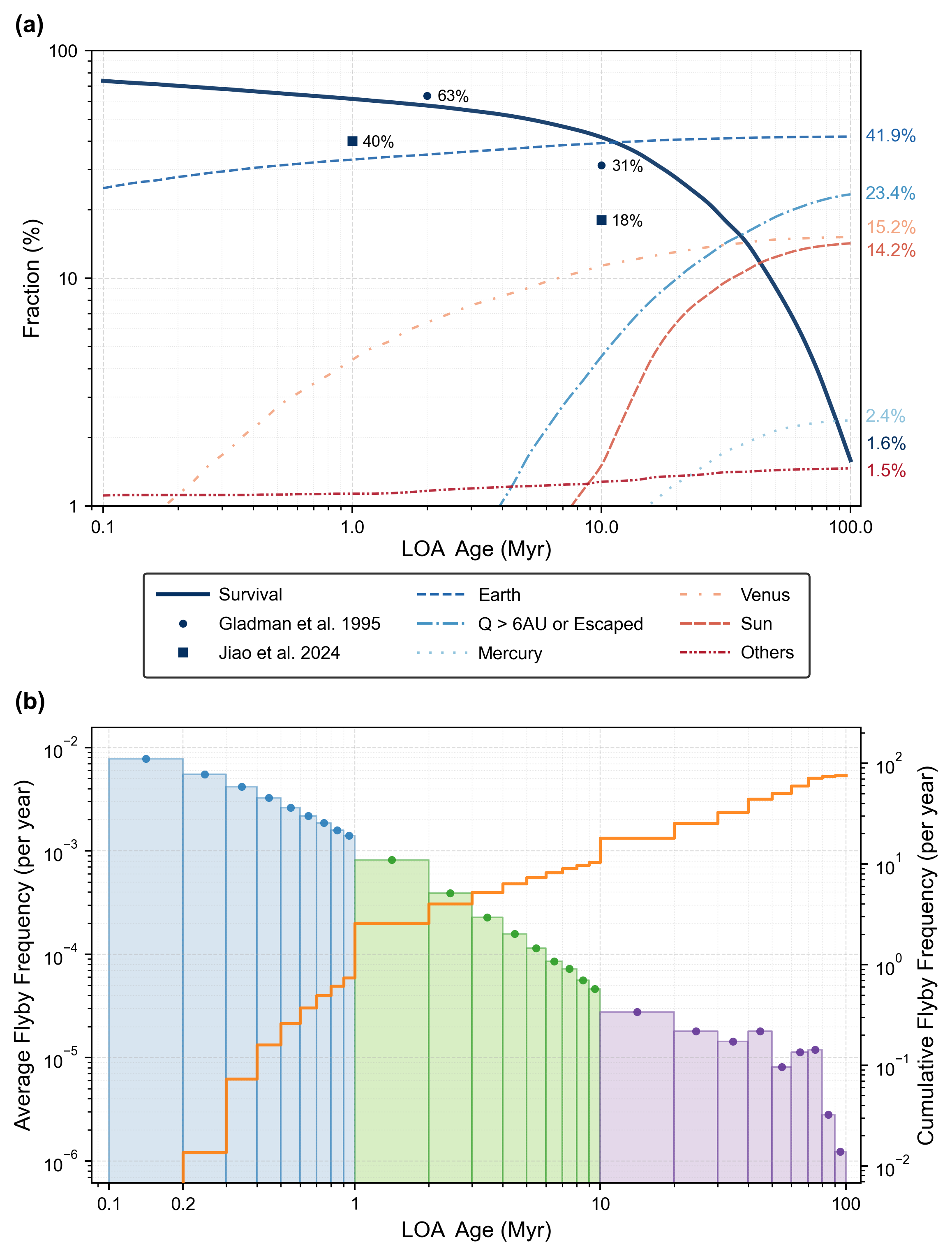} 
\caption{\textbf{Dynamical fates and flyby frequency of globally sourced lunar ejecta.} (a) illustrates the temporal evolution of 20,000 test particles, which simulate lunar ejecta launched from the entire lunar surface. Presented on a log-log scale, the colored curves show the fraction of these particles in various dynamical states: survival, impact (with specific planets or the Sun), reaching an aphelion of $Q > 6$~AU, and others. The final fraction for each state at 100~Myr is annotated on the right side of the plot. For comparison, the survival fractions reported by \citet{jiao2024asteroid} and \citet{gladman1995dynamical} are also plotted. (b) shows the present-day LOA flyby frequency. The bars (left \textit{y}-axis) present $\lambda(t)$, the average flyby frequency per initial ejecta for LOAs of that age; the curve (right \textit{y}-axis) is the production‑rate‑weighted cumulative number of annual LOA flybys up to that age, equal to $N_\text{flyby}$ at 100~Myr.
\label{fig:2}}
\end{figure*}

\subsection{Dynamical fates of the lunar ejecta}\label{sec:3.1}

Our N-body simulations show that the depletion of LOAs accelerates with age. 
As shown in Figure~\fref{fig:2}{a}, the survival fraction is 74\% for a 0.1-Myr-old population, declining to 61\% at 1~Myr and 42\% at 10~Myr, before plummeting to a mere 1.6\% for a 100-Myr-old population. 
Our LOA survival fraction is consistent with that of \citet{gladman1995dynamical} at early times (~63\% at 2~Myr), but higher by a factor of ~2 after 10~Myr than that reported by \citet{jiao2024asteroid} (18\% at 10~Myr and only 0.7\% at 100~Myr). 
The lower fraction from the \citet{jiao2024asteroid} study is due to that it is limited to a single source crater and does not account for the Yarkovsky effect, in contrast to our more comprehensive model.

Our results also show that within the first 0.1~Myr, 24.8\% of ejected particles hit Earth (Figure~\fref{fig:2}{a}). 
This aligns well with \citet{castro2025lunar} and \citet{gladman1995dynamical}, who respectively reported that 22.6\% and 20$\sim$25\% of lunar ejecta would collide with Earth in the same timescale. 
Furthermore, it's indicated that lunar ejecta impact Earth predominantly within the first 1~Myr, specifically, with the most occurring within 0.1~Myr. 
This timescale is consistent with the typical lunar meteorite transit timescale from Moon to Earth ($< $ 1~Myr, often tens of thousands of years) inferred from cosmic ray exposure ages \citep{joy2023lunar,hidaka2017isotopic}. 

Notably, we find that after an evolution of 10--100~Myr, the proportion of lunar ejecta impacting Earth is 41.9\%, which is lower than the 60\% reported by \citet{jiao2024asteroid} and 56\% by \citet{gladman1995dynamical} at 10 Myr.
Meanwhile, the final proportions (at 100~Myr) of ejecta leaving the Sun for 6~AU, falling onto the Sun, and impacting Mercury are 23.4\%, 14.2\%, and 2.4\%, respectively. 
This suggests that on timescales of 10-100 Myr, long-term dynamical evolution shifts the ultimate fate of these objects away from Earth impact and toward other removal channels, which is likely enhanced by the Yarkovsky effect.

\subsection{The current lunar-NEA population }
 
Based on the scaling law in Section~\ref{sec:2.1}, the flux of the lunar escaped ejecta can be obtained integrally as
\begin{align}
N_{\text{e}}(>D,<t)=\int_{D_\text{{c,min}}}^{D_\text{{c,max}}(t)}kD^{-p}D_{\text{i}}^3\cdot bN_0(t)D_{\text{c}}^{-b-1}{\rm d}D_{\text{c}}~.
\end{align}
The plausible range of the maximum crater diameter, $D_\text{{c,max}}(t)$, is constrained by the ``background'' model and ``real'' model presented in Section~\ref{sec:2.1}, and the minimum one $D_\text{{c,min}}$ can be constrained by the minimum impactor diameter $D_\text{{i,min}}$ for target ejecta (as indicated by the red vertical line in Figure~\ref{fig:1}). From this, we define the age-dependent production rate as
\begin{align}
&Q(t) = N_{\text{e}}(>D,<t)/t~.
\end{align}
For example, the ``background" model indicates a production rate of ~63,000 LOAs ($D > 5$ m) per Myr for the 100-Myr-old population.

By combining these estimated production rates with the survival fraction $S(t)$ for lunar ejecta of  different ages (Figure~\fref{fig:2}{a}), we can estimate the current LOA population as
\begin{align}
&N_{\text{LOA}}(>5~m)=\int_{t_{\text{c,min}}}^{t_\text{{c,max}}} S(t)Q(t){\rm d}t~,
\end{align}
The minimum crater age $t_\text{c,min}$ corresponds to the condition $D_\text{{c,min}} = D_\text{{c,max}}(t)$, yielding $t_\text{c,min} > 0.1$ Myr for both models. The maximum age $t_\text{c,max}$ is limited to 100~Myr, because the lunar ejecta older than this age shows a negligible survival fraction ($<$~1.6\%).
Finally, our ``background'' model predicts that approximately 260,000 LOAs ($D > 5 \text{\ m}$) are still present in near-Earth space, while our ``real'' model suggests a population of about 500,000. Even with poor survival rates (42\%$\sim$1.6\%) for ejecta beyond 10~Myr, the integrated contribution of craters older than 10~Myr accounts for over 90\% of the present-day LOAs.

Surviving LOAs are rare, constituting only $\lesssim$ 1\% of the total NEA population ($D > 5 \text{\ m}$, estimated to be $10^8$ objects, \citealt{bottke2015collisional}). However, their close flybys of Earth offer a key opportunity for LOA discoveries, and each new detection serves to validate this theoretically predicted population.

\begin{figure*}[ht!]
    \centering
    \includegraphics[width=0.9\linewidth]{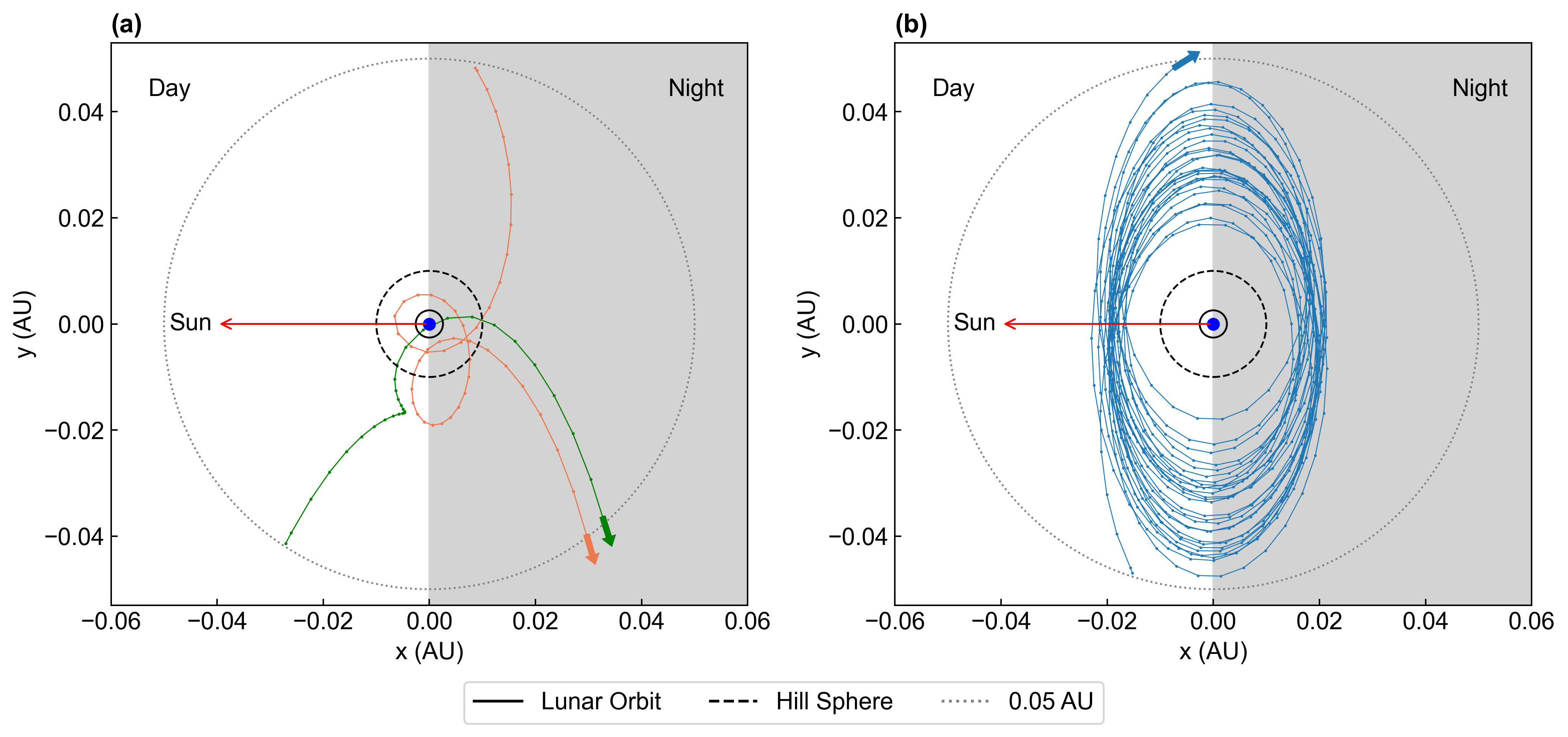} 
\caption{\textbf{Illustrative distinct modes of LOA Earth flybys.} The trajectories of LOAs passing within 0.05 AU of Earth are projected onto the ecliptic (\textit{x-y}) plane in the geocentric Sun\textendash Earth co-rotating frame. An arrow at the end of each trajectory indicates the particle’s direction of motion; the positive \textit{y}-axis points toward Earth's leading side, and the red arrow points toward the Sun. (a) shows two low-energy capture modes: the orange trajectory represents a `minimoon' state (e.g., 2024~PT5), characterized by a close passage ($<$~0.01~AU) and negative geocentric Keplerian energy. The green trajectory depicts a close flyby, where the object slows to a near-zero velocity with a significant change in direction. (b) illustrates a quasi-satellite state (e.g., Kamo‘oalewa), where the object closely co-orbits Earth for decades.
\label{fig:3}}
\end{figure*}

\subsection{The Earth flybys of LOAs }

In this work, an Earth flyby is defined as an event in which a heliocentric particle passes within 0.05~AU of Earth, consistent with the minimum orbital intersection distance (MOID) threshold for potentially hazardous asteroids (PHAs) \citep{perna2015grasping}.
For our analysis, we focus on particles that remain within this proximity for at least 0.15 years (5 output steps). 

Among the numerous Earth flyby events in our simulations, several particularly interesting modes are identified. 
In Figure~\fref{fig:3}{a}, the orange trajectory shows a temporary capture ($\lesssim$ 4 months) by the Earth\textendash Moon system, similar to the so-called `minimoon' 2024~PT5 (passing inside Earth's Hill sphere with a temporary negative two-body energy, \citealt{kareta2025lunar}).
The green trajectory shows the particle's velocity approaching zero (in the Sun\textendash Earth rotation frame) at a point outside the Hill sphere, with a significant change in its direction that results in a close Earth flyby. 
Differing from these transient interactions, Figure~\fref{fig:3}{b} shows a quasi-satellite orbit that lasts for decades within 0.05~AU of Earth, which is similar to that of 2016~HO3 but much closer to Earth. 

To quantify the present-day LOA flybys, we first calculate $\lambda(t)$, the average flyby frequency per initial ejecta ($D > 5 \text{\ m}$) in an LOA population of a given age $t$ (0.1--100~Myr). 
This implicitly accounts for the population's dynamical decay.
The total annual number of LOA flybys is then calculated by integrating $\lambda(t)$ weighted by the production rate $Q(t)$, as
\begin{align}
N_{\text{flyby}}=\int_{t_{\text{c,min}}}^{t_\text{{c,max}}} \lambda(t)Q(t){\rm d}t~.\label{eq:12}
\end{align}
This calculation yields an annual rate of approximately 40 LOA flybys ($<0.05$~AU) for the ``background'' model, and 76 for the ``real'' model (Figure~\fref{fig:2}{b}). 
Furthermore, the Earth flybys of LOAs are dominated by older craters, with LOAs older than 10~Myr contributing the most flybys (35 of 40 for ``background'' model and 65 of 76 for ``real'' model). 

As a comparison, the population of PHAs\textemdash NEAs with diameters greater than 140 m that can approach Earth within 0.05~AU over the next century\textemdash is 4700$\pm$1450 \citep{perna2015grasping}. 
According to \citet{bottke2015collisional}, the approximate number of NEAs is $10^6$ for $D > 140 \text{\ m}$ and $10^8$ for $D > 5 \text{\ m}$. 
Consequently, there are about 5,000 annual Earth flybys by NEAs ($D > 5 \text{\ m}$) at present, with LOAs potentially accounting for 1$\sim$2\% of these encounters.
Despite these rough estimates, approximately 1,600 NEA flybys within 0.05~AU of Earth were observed in 2024\footnote{\url{https://cneos.jpl.nasa.gov/ca/}}. 
Considering our estimated LOA population and their Earth flyby frequency relative to NEAs ($\lesssim$~2\%), we would expect 10--20 LOAs in these annual observations, which will be further examined in the next section.

\begin{figure*}[ht!]
    \centering
    \includegraphics[width=\linewidth]{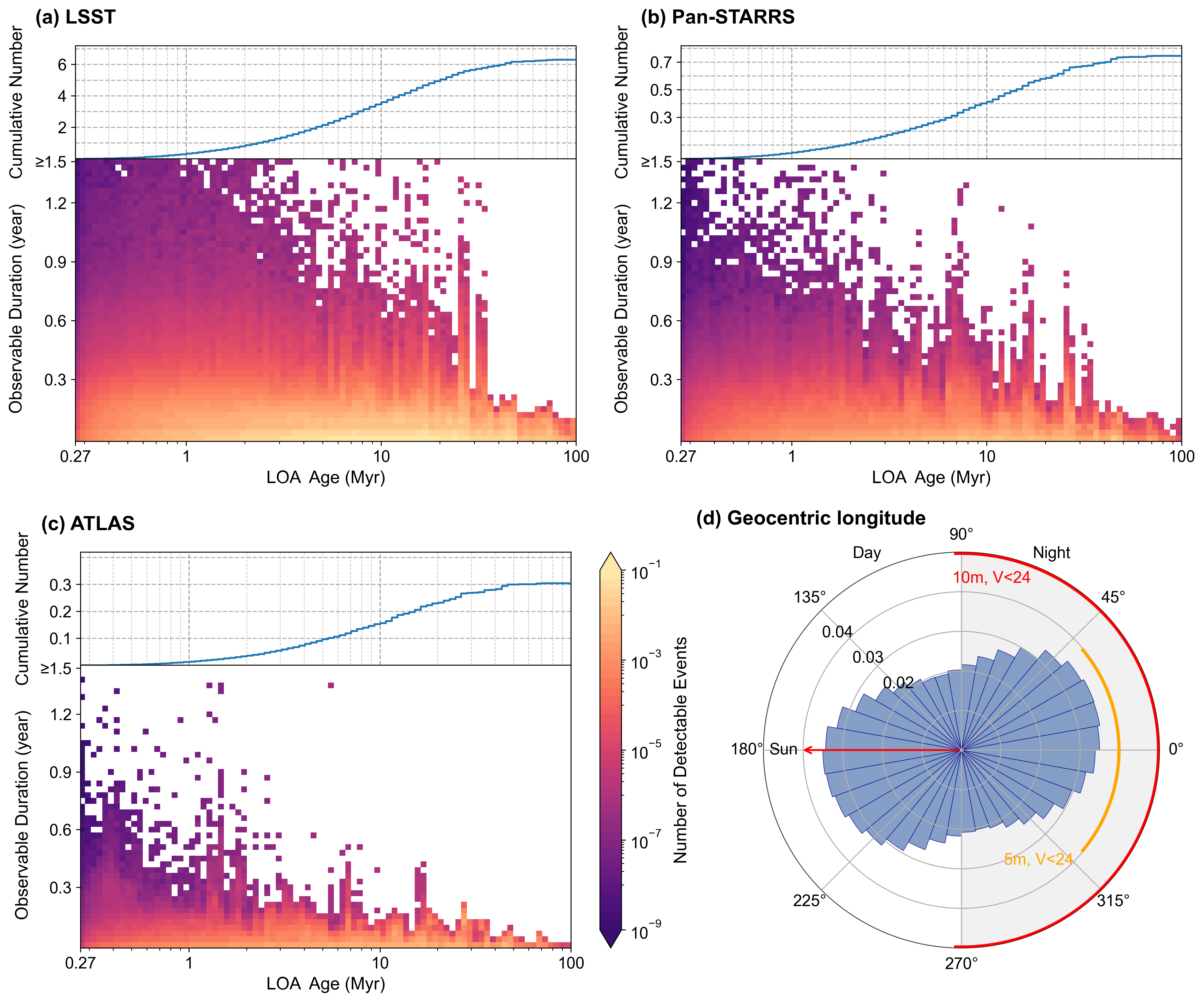} 
\caption{\textbf{Detection efficiency distribution for LOA flybys.} (a--c) show heatmaps of the annual number of LOA detections by LSST (a), Pan-STARRS (b) and ATLAS (c),binned by asteroid age and the observable duration. We apply survey-specific detection limits: LSST ($V_\text{trail}<24$, $\omega<10~\text{deg day}^{-1}$), Pan-STARRS ($V_\text{trail}<22$, $\omega<5~\text{deg day}^{-1}$), and ATLAS ($V_\text{trail}<20$, $\omega<10~\text{deg day}^{-1}$). Above each heatmap is the cumulative age-distribution curve, equal to $N_\text{det}$ at 100~Myr. (d) shows the relative frequency distribution of geocentric ecliptic longitude for LOA flybys at 0.05~AU from Earth, where 0$^{\circ}$ marks the anti-solar direction and 90$^{\circ}$ the direction of Earth's motion. The red and orange arcs indicate the detectable range ($V<24$) for asteroids with diameters of $D=10~\text{m}$ and $D=5~\text{m}$, respectively.
  \label{fig:4}}
\end{figure*}

\subsection{Detection efficiency for LOAs}

The challenge in detecting LOA flybys lies in two key factors: their apparent magnitude $V$ and their orbital angular velocity $\omega$ relative to Earth. Fast-moving objects appear as streaks in telescopic images, a phenomenon known as "trailing loss" \citep{deienno2025debiased}, which effectively lowers their measured brightness. We therefore account for trailing by adopting the trailed magnitude $V_\text{trail}$, defined as
\begin{align}
&V_\text{trail}=V+\Delta V(\omega)~.
\end{align}
Here, $\Delta V(\omega)$ is specific to each survey's characteristic exposure time and Full Width at Half Maximum (FWHM).

The upcoming Legacy Survey of Time and Space (LSST) is uniquely equipped to tackle this challenge, notably with a limiting magnitude of approximately 24. Using the formulation described in \citet{nesvorny2024neomod}, \citet{jedicke2025steady} calculated the $\Delta V(\omega)$ for LSST as
\begin{align}
\Delta V(\omega)= 2.5 \log_{10}{[1 + 1.25(\omega - 0.8)]}~.\label{eq:14}
\end{align}
Furthermore, an angular velocity limit of 10~deg day$^{-1}$ is applied, as it is imposed by the LSST to distinguish natural objects from artificial ones \citep{jedicke2025steady}. 
We also adopt a minimum 20-day observable window to allow follow-up confirmation.
Thus, we can estimate the current annual LOA detections by LSST using an integral analogous to that for the LOA flyby frequency as
\begin{align}
N_{\text{det}}=\int_{t_{\text{c,min}}}^{t_\text{{c,max}}} \lambda_{\text{det}}(t)Q(t){\rm d}t~.\label{eq:15}
\end{align}
Here, $\lambda_{\text{det}}(t)$ is the detectable flyby frequency obtained from $\lambda(t)$ in Equation~\eqref{eq:12} after incorporating the observational constraints above.
Applying the expected LSST detection efficiency to our models, we predict that the survey will discover approximately 2.9 and 6.3 LOAs ($D > 5 \text{\ m}$) per year for the ``background'' model and ``real'' model, respectively, which is a considerable number. 

As shown in Figure~\fref{fig:4}{a}, LOAs potentially detectable during Earth flybys are almost exclusively older than 1 Myr, with roughly equal numbers originating from populations younger and older than 10 Myr. Although LOAs older than 50 Myr contribute nearly half of total Earth flybys, they produce very few detectable events, indicating that most flybys of LOAs older than 50 Myr occur at higher encounter velocities and are therefore missed by ground-based surveys.

In addition to LSST, we also assess the LOA detection efficiency for the Panoramic Survey Telescope and Rapid Response System (Pan-STARRS) and the Asteroid Terrestrial-impact Last Alert System (ATLAS) under the ``real'' model.
For Pan-STARRS, we use a limiting magnitude of 22 and a maximum detectable angular velocity of 5 deg/day \citep{denneau2013pan}. 
Given its exposure mode and FWHM similar to those of LSST \citep{chambers2016pan}, we again use Equation \eqref{eq:14} to estimate its $V_\text{trail}$.
For ATLAS, it can detect trail intensities fainter than 20.5 near 10~deg day$^{-1}$ \citep{heinze2021neo}, with its trailing loss $\Delta V(\omega)$ given by \citet{deienno2025debiased} as
\begin{align}
\Delta V(\omega)= 2.5 \log_{10}{(0.501\sqrt{\omega})}~.
\end{align}
Furthermore, ATLAS's twice-nightly sky coverage \citep{tonry2010early} justifies our use of a shorter duration limit ($>$ 10 days) for it.
Using the survey-specific limits above, we apply Equation \eqref{eq:15} to calculate the cumulative detectable events for LOAs by Pan-STARRS (Figure~\fref{fig:4}{b}) and ATLAS (Figure~\fref{fig:4}{c}), yielding an average of 0.75 LOA discoveries per year by Pan-STARRS and 0.31 by ATLAS.
These much lower detection rates, compared to the LSST, directly explain the current scarcity of LOA candidates and highlight the transformative potential of the next generation of surveys.

\begin{figure*}[ht!]
    \centering
    \includegraphics[width=0.65\linewidth]{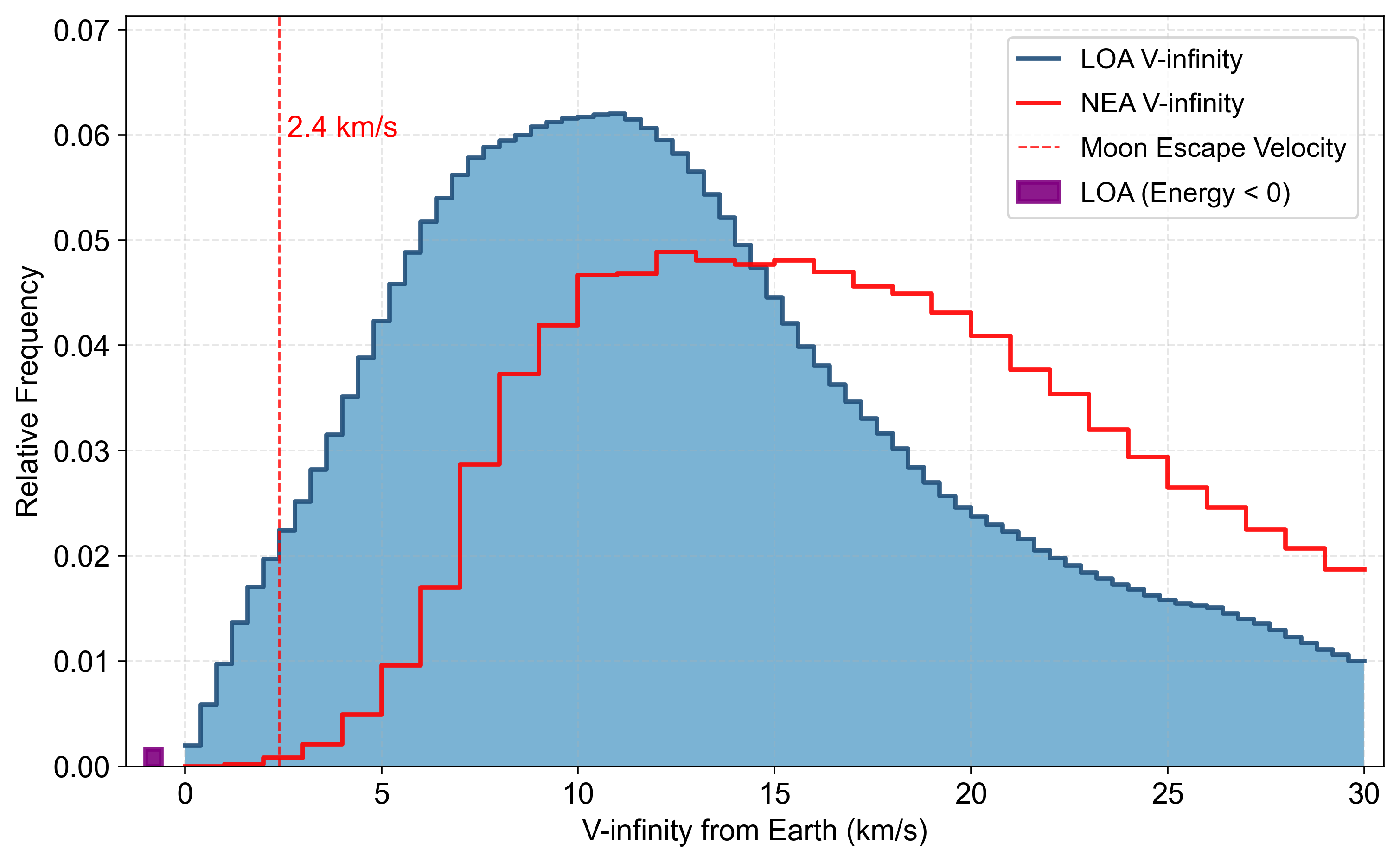} 
\caption{\textbf{Distributions of encounter velocity.} Here we present the relative frequency distributions of LOAs' and NEA's velocity at infinity ($v_{\infty}$, with a dedicated bin for LOAs with a negative geocentric Keplerian energy.
\label{fig:5}}
\end{figure*}

\subsection{Dynamical characteristics of LOAs}

While spectroscopy can offer definitive proof of lunar origin \citep{kareta2025lunar}, it is not practical to survey the vast NEA population. 
A targeted search strategy is therefore essential to efficiently identify more LOA candidates for spectroscopic follow-up. 
Here, we present two key dynamical characteristics of LOA flybys that provide clear guidance for such efforts.

\subsubsection{Lower Earth-flyby velocities}

The velocity at infinity ($v_\infty$) is a key measure of an object's geocentric energy and is quasi-conserved in the Sun\textendash Earth CRTBP system \citep{murray1999solar,carusi1995conservation}, making it a robust indicator of Earth flyby dynamics. 
Analyzing the velocity distribution of all LOA flyby events within 0.05 AU, weighted by their age-dependent production rate $Q(t)$ under the ``real'' model, we find a median of 11.9~km/s and a mean of 12.8~km/s (Figure~\ref{fig:5}). 
In comparison, \citet{heinze2021neo} provide the NEA encounter velocity ($\approx v_\infty$) distribution within 0.05~AU (the red line in Figure~\ref{fig:5}), which has a median of 17.5~km/s; a separate 100-Myr simulation places the average velocity at Earth's Hill sphere at 22.5~km/s \citep{ito2010asymmetric}. 
The LOA $v_\infty$ distribution exhibits a sharper peak and is significantly skewed toward lower velocities compared to that of NEAs, highlighting the low-velocity nature of LOA flybys likely due to their origin in the Earth\textendash Moon system.
This slower encounter velocity generally corresponds to a lower angular velocity \citep{heinze2021neo}, potentially increasing the detection efficiency for LOAs.

Notably, the probability for an asteroid to have a low Earth encounter velocity, specifically below the lunar escape velocity ($v_{\infty} < 2.4 \text{ km/s}$) is dramatically higher for LOAs than for the general NEA population: 2.8\% versus 0.05\%. 
As a result, despite LOAs comprising only $\lesssim$~1\% of all NEAs, we estimate that a newly discovered NEA with $v_{\infty} < 2.4 \text{ km/s}$ may have a probability as high as ~30\% of being of lunar origin. 
Our work thus quantifies the suggestion by \citet{kareta2025lunar}. 
We find that while a low $v_{\infty}$ is not definitive proof, its strong correlation with lunar origin establishes it as a powerful indicator for identifying promising candidates.

Given that the recently discovered candidate LOA 2024 PT5 is a minimoon \citep{kareta2025lunar}, it is worth examining the capture probability for LOAs. 
Counting flybys with a negative geocentric Keplerian energy, we find that they account for only about 0.07\% of all flybys. 
Thus, while LOAs ($D > 1 \text{\ m}$) may constitute a steady-state population of Earth's minimoons \citep{jedicke2025steady}, temporary capture events represent only a minor fraction of all LOA encounters with Earth. 
The dominant source of future LOA discoveries will therefore be the more numerous passing-through flybys.

\subsubsection{A Biased Approach Geometry}

The approach direction distribution of LOAs is critical to understanding their detectable characteristics, with particular relevance to early warning for planetary defense. 
We analyze the distribution of geocentric ecliptic coordinates (longitude and latitude) for all LOA flybys entering Earth's Hill sphere (weighted by $Q(t)$), whose radius is 0.05 AU in the Sun\textendash Earth co-rotating frame.

The flyby distribution exhibits an almost linear decrease with increasing absolute latitude, with very few objects observed beyond 80$^{\circ}$. Surveys for LOAs should therefore prioritize low-latitude regions.
Regarding longitude (Figure~\fref{fig:4}{d}), the LOA flybys are strongly concentrated from $-20^{\circ}$ to 50$^{\circ}$ and from 160$^{\circ}$ to 230$^{\circ}$, and these regions are symmetric.
Since a significant portion of LOAs have low energy during Earth flyby, they are restricted to approaching from sunward and anti-sunward directions, rather than from the high-potential-energy regions (or forbidden regions) leading and trailing Earth in the Sun-Earth co-rotating frame.
Furthermore, the longitude distribution shows a slight counterclockwise tilt relative to the Sun\textendash Earth axis, a feature likely driven by the transport properties of invariant manifolds in the CRTBP \citep{gomez2004connecting,swenson2019invariant,dermawan2025dynamical}. 

We also calculated the apparent magnitudes for $D=10 \text{ m}$ and $D=5 \text{ m}$ asteroids at 0.05~AU, overlaying the resulting detecability arcs where $V<24$ in Figure~\fref{fig:4}{d}. 
While most LOA flybys from the anti-sunward hemisphere can be detected at this distance, sunward-approaching LOAs remain largely undetectable by ground-based surveys, creating a significant blind region for planetary defense.

\section{Conclusion}

In this work, we trace the 100-Myr orbital dynamics of the lunar ejecta from the entire lunar surface by coupling a production model with an N-body simulation. 
Our results suggest:

\begin{itemize}[topsep=0em, itemsep=0em, leftmargin=1.5em]
    \item There are estimated 500,000 extant LOAs with diameter $>$~5~m, which contribute $\lesssim$~1\% to the population of NEAs of the same size.
    \item Annually, an estimated average of over 70 LOAs ($D > 5 \text{\ m}$) fly by within 0.05~AU of Earth, accounting for $\lesssim$~2\% of flyby events of typical NEAs of the same size.
    \item Considering the trailed apparent and angular velocity, LSST is expected to observe an average of 6 LOA flybys annually, suggesting a promising future for LOA discovery.
    \item LOAs exhibit significantly lower encounter velocities with Earth compared to NEAs; objects with $V_{\infty} < 2.4 \text{ km/s}$ demonstrate a high probability (30\%) of lunar origin.
    \item The LOA approach distribution at 0.05~AU peaks at the ecliptic plane; in longitude, it is strongly concentrated in two symmetric ranges: $-20^{\circ}$ to 50$^{\circ}$ and $160^{\circ}$ to $230^{\circ}$.
\end{itemize}

As a summary, our results build a complete picture of the dynamical journey of LOAs, from their origin from lunar impacts to their appearance in Earth's vicinity, providing a quantitative forecast for their discovery in the LSST era.
A forthcoming paper (in preparation) will apply this framework to the impending 2024~YR4 impact to predict the resulting new LOA population and its associated flux of lunar meteorites at Earth.

\begin{acknowledgments}
This work is supported by the National Natural Science Foundation of China under Grant Nos. 62227901, 125B1015, 123B2038, U24B2048, 12572404, 12372047, the national level fund No. KJSP2023020301, the Beijing Nova Program No. 20250484831, and the Tsinghua University Initiative Scientific Research Program (Student Academic Research Advancement Program).

\end{acknowledgments}

\begin{contribution}
YJ and BC initiated the conception of this project.
YW performed the simulations, analyzed the numerical results, and led the writing of the manuscript.
YJ, WD, YH and BC contributed to the interpretation of the results and the writing and revision of the manuscript.
ZL contributed to the interpretation of the result.
JL and HB contributed to the writing and revision of the manuscript.
\end{contribution}




\bibliography{sample7}{}
\bibliographystyle{aasjournalv7}



\end{document}